\documentclass{llncs}
\usepackage{graphicx}
\usepackage{pbox}
\begin{document}

\bibliographystyle{splncs}

\title{An exploratory study on how Internet of Things developing companies handle User Experience Requirements}
\author{Johanna Bergman\inst{1} \and Thomas Olsson\inst{2} \and Isabelle Johansson\inst{1} \and Kirsten Rassmus-Gr\"ohn\inst{1}}
\institute{Department of Design Sciences, Lund University \email{johanna.e.bergman@gmail.com, kirsten.rassmus-grohn@certec.lth.se, isabelle.a.e.johansson@gmail.com} \and RISE SICS AB \email{thomas.olsson@ri.se}}
\subtitle{This is a pre-print of paper accepted at the 24th International working conference on Requirements Engineering: Foundation for Software Quality, 19-22 March, 2018, Utrecht, The Netherlands}
\maketitle
\begin{abstract}

\textbf{[Context and motivation]} Internet of Things (IoT) is becoming common throughout everyday lives. However, the interaction is often different from when using e.g. computers and other smart devices. Furthermore, an IoT device is often dependent on several other systems, heavily impacting the user experience (UX). Finally, the domain is changing rapidly and is driven by technological innovation. 

\textbf{[Question/problem]} In this qualitative study, we explore how companies elicit UX requirements in the context of IoT. A key part of contemporary IoT development is also data-driven approaches. Thus, these are also considered in the study. 

\textbf{[Principal idea / Results]} 
There is a knowledge gap around data-driven methodologies, there are examples of companies that collect large amount of data but do not always know how to utilize it. Furthermore, many of the companies struggle to handle the larger system context, where their products and the UX they control are only one part of the complete IoT ecosystem. 

\textbf{[Contribution]} We provide qualitative empirical data from IoT developing companies. Based on our findings, we identify challenges for the companies and areas for future work. 
\end{abstract}

\section{Introduction}\label{sec:intro} 
Internet of Things (IoT) is rapidly growing and will have a fundamental impact on our lives. IoT is advancing into many domains, facing new contexts and usages, such as hospitals, smart buildings, wearables and smart vehicles. The interaction  with IoT is often different than for e.g. a computer or smart phone~\cite{DeHaan2015}.

The nature of IoT extends the interaction possibilities through mobile and wireless networks, social and collaborative applications, connected data, and the use of intelligent agents \cite{DeHaan2015}. The diverse nature of interaction possibilities with IoT results in that the product being developed will be part of a whole ecosystem of devices \cite{Fauquex2016}. Furthermore, the combination of hardware and software design is a distinguishing part of the design methodology for IoT \cite{Rowland2015}. IoT affects the design methodology and processes through increased importance of the user-centeredness of design where the user actively can determine the design outcome, increased use of higher level tools and applying new, agile, and exploratory design methods \cite{DeHaan2015}. At the same time, innovation and deciding what to implement is more customer-driven and based on data from actual usage \cite{bosch_speed_2016}. 

The term User Experience (UX) can be defined as \textit{“a person's perceptions and responses resulting from the use and/or anticipated use of a product, system or service”} \cite{iso9241}. As such, UX attempts at capturing all aspects of the experience of using a product, system or service, such as emotions and perceptions in all stages of use, the perception of brand image, performance, and the context of use. Similar to usability \textit{“the extent to which a system, product or service can be used by specified users to achieve specified goals with effectiveness, efficiency and satisfaction in a specified context of use”} \cite{iso9241}, UX is typically considered to be a quality requirement (QR) or non-functional requirement (NFR) \cite{glinz2007nfr,iso25010}. However, UX is inherently difficult to measure, while usability can be measured objectively, e.g. time to complete tasks, and subjectively, e.g. the system usability scale (SUS) \cite{brooke1996sus}. Usability can furthermore be seen as a subpart of UX, which underlines the attempt of UX at capturing universal and overall qualities of an individual using a product, system or service. Fraifer et al. proposes a quantifiable way of communicating and describing UX based on 84 different (mainly subjective) evaluation methods such as hedonic qualities, diary studies, interview and questionnaire guides, experience sampling, etc.~\cite{Fraifer2017}. They create a radar diagram based on the overarching qualities \textit{Look} (Visual Design, Information structure, Branding), \textit{Feel} (Mastery of interaction, Satisfaction, Emotional Attachment) and \textit{Usability}. The concept of UX also touches on the meaning it creates in a user’s life, and what needs it fulfills~\cite{Hassenzahl2013,Hassenzahl2015}. Hassenzahl comments that even though the concept of UX (in his words: \textit{proposition to consider the experience before the thing}), has been adopted by academics and HCI practitioners, not much has changed in the general design approach \cite{Hassenzahl2013}.

In this paper, we study how IoT development companies address IoT UX. This is part of our ongoing efforts to understand the overall decision process around IoT system development. Specifically, this paper aims at understanding the activities performed in the context of data-driven development to decide how to address UX requirements. We define the following research questions: 
\begin{itemize} 
\item [ ]  
\begin{enumerate}
      \item [RQ1] How are UX requirements elicited in the context of IoT development in general?  
      \item [RQ2] How are data-driven methodologies specifically utilized for IoT development to elicit UX requirements?  
      \item [RQ3] Which are the challenges for UX and IoT?
  \end{enumerate}
\end{itemize}

This paper is organized as follows: In Section~\ref{sec:rw}, the related work is outlined. The research method used is described in Section~\ref{sec:method}. Section~\ref{sec:results} presents the main results and Section~\ref{sec:discussion} summarizes the discussion. The paper is concluded in Section~\ref{sec:conclusion}. 

\section{Related Work}\label{sec:rw} 
IoT interfaces pose certain challenges, in that a large part of the interaction going on is invisible to the user (ubiquitous cf. \cite{resnick2013ubiquitous}). Furthermore, the technology is, in itself, distributed and asynchronous and each IoT device typically consists of a combination of a physical product, underlying software and network services \cite{Rowland2015}. This affect the way the user is able to interact with it. IoT therefore impacts the design process, putting a larger focus on UX evaluation and design methods that can enhance UX \cite{DeHaan2015,Fauquex2016,Kranz2010}. This is accomplished by, for example, using agile \cite{warden2007art} development, iterative design and prototyping, and applying user-centered design principles (e.g. \cite{iso9241}), rather than traditional requirements engineering.

One way of accomplishing an iterative design, that meets the users’ needs and expectations, is to improve the system continuously after it has been released to the market, for example by collecting usage data (analytics) \cite{Rowland2015}. However, the physical design of the device is less flexible, and changing the physical product after launch is typically never performed and entails large costs. Therefore, iteration and parallel design, and conceptualizing the product in the design process become more important for the hardware part. Lin et al. attempt at creating a framework for how to combine the data-driven approach with product form design \cite{Lin2016}. The main part of the framework consists of conducting a UX scenario experiment with the product. However, they conclude that the limitation of working iteratively with the physical object results in that the presented framework can mainly be used by newly launched products with short life cycles. 

According to Pallot et al. \cite{Pallot2015}, there is in general more research conducted on UX evaluation (subjective) rather than UX measurement (objective). Furthermore, they consider that, due to the complexity of UX, most papers in the field describe a narrow UX evaluation, focusing on ergonomic and hedonic qualities. In the context of an experiential living lab for IoT, they have elaborated on the UX life-cycle described by Roto et al \cite{roto2011user}, and proposed a UX framework and model with a combination of 42 different properties in three categories: Knowledge, Social and Business. 

In addition to hedonic and ergonomic qualities, they single out three of the Business UX elements specifically concerning IoT; automation level, connectivity, and reliability. However, their ideas of how to conduct measurements per se are not elaborated.

One large part of consumer IoT is wearables. The increasing use of wearables is referred to by Barricelli et al. \cite{Barricelli2017}, as the \textit{quantified-self} movement. In \cite{Oh2015}, Oh and Lee discuss UX issues for quantified-self. It is stated that the wearables are often regarded as fashion items and therefore aesthetics is important. The size and shape are also said to play a role in order not to disturb the user. Shin investigates the term quality of experience (QoE), that he describes as encompassing both UX and the quality of service (QoS), and shows how they are interrelated~\cite{Shin2017}. However, Shin does not define what they mean by UX, but describes it to be related to usefulness and enjoyment. 

Ovad and Larsen conducted a study on UX and usability in eight different Danish companies \cite{Ovad2015}. Their mainly focused on how to combine agile development methods with UX. Three were software companies and five were companies working with embedded software in physical products. They argue that there is a gap between industry and academy when it comes to UX and usability methods. Holmstr\"{o}m-Olsson et al. studied five different Swedish companies’ view on interaction and ecosystems for IoT \cite{Olsson2016}. Their study also presents a model (User Dimensions in IoT, UDIT), that is focused on user interaction rather than a broader view of UX.

Customer-driven innovation and a close communication with the users is an important trend in software engineering. Customer-driven understanding means understanding the specific and detailed needs of the customers as a vehicle for innovation rather than being technology driven \cite{bosch_speed_2016}. There is also a movement to work with concrete data rather than informed opinions \cite{Problem2014}. This is closely related to working with continuous deployment and creating an atmosphere where the users are used to "being experimented on"~\cite{parnin2017top}.
With the study presented in this article, the existing work is complemented with how IoT development companies actually use data and analytics to understand the UX requirements, both for hardware and software.

\section{Research Method}\label{sec:method} 
\begin{figure}[!hbt]
\centering
\includegraphics[scale=0.45]{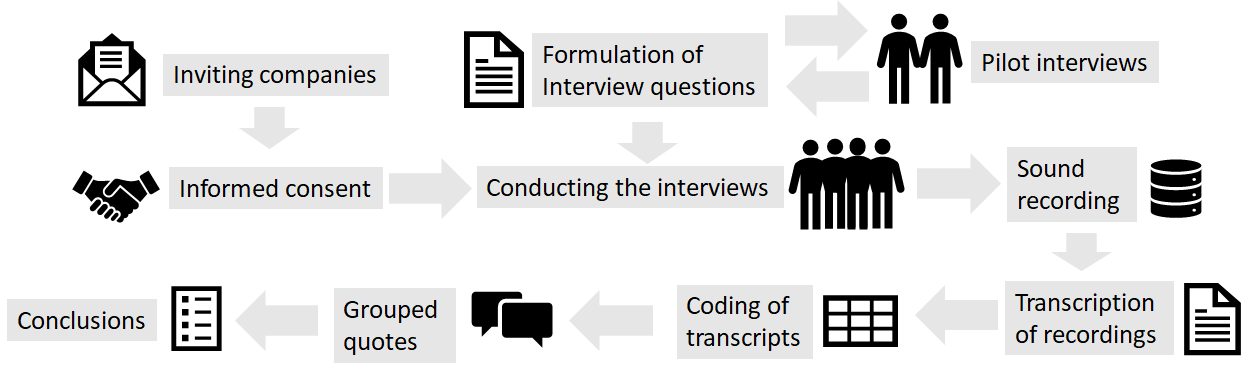} 
\caption{Overview of the method for the exploratory case study.}
\label{fig:empirical}
\end{figure}

Considering the exploratory nature of this study and the aim to describe the diversities among companies within the defined area, the qualitative approach was found to be the most suitable \cite{easterbrook2008}. The overall design of the study is found in figure \ref{fig:empirical}.

\subsection{Data Collection}
Data was collected through semi-structured interviews \cite{runeson2012}. The interview instrument was structured according to the funnel model meaning that the character of the questions moved from general to specific \cite{runeson2012}. The instrument was evaluated in two pilot interviews, resulting in some adjustments. The interview instrument used can be found in \cite{johannaisabelle2017}. 

\subsection{Execution}
The analysis was qualitative with the aim to explore and gain understanding; not to explain and statistically analyze. The selection of companies was based on a combination of convince sampling and maximum variation \cite{flyvbjerg} (cf. \cite{johannaisabelle2017} for details). The participating companies all develop IoT products or systems. All companies have an office in the south part of Sweden, and all interviews except from one took place at their respective office. The companies and interviewees' roles are summarized in Table~\ref{tab:total}. Companies A-C and E-G are consultancy companies, where Company G is a design studio and the others cover the complete process. Companies D and H-K are product development companies. Software is central to all of the companies and they develop complete software systems. That is, they do not merely develop the software embedded in their hardware products. The older companies (A, E, F and H) come from other domains and have over time started working with IoT. 

The interviews were performed by the first and third author. All the interviews were recorded and lasted for approximately one hour. Both interviewers asked questions and interacted in with the interviewees. In two cases (E and C), there were two interviewees. The interviewees were selected based on their insights into the requirements and UX processes. 

\begin{table}[!hbt]
\centering
\includegraphics[scale=0.9]{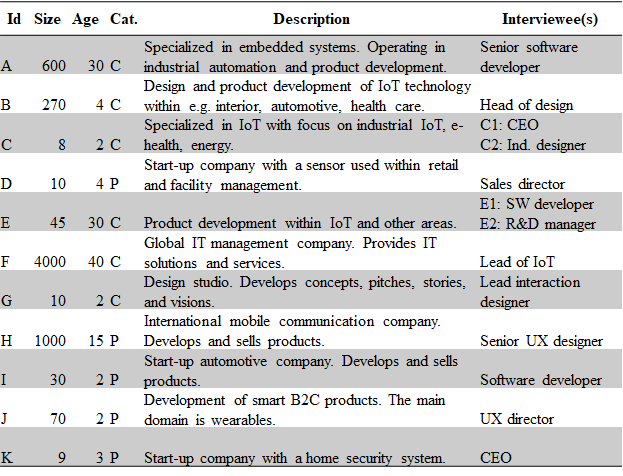} 
\caption{Participating companies. Category (Cat.): P = Product company, C = Consulting company. The sizes of the companies are displayed as number of employees, and the age in years.}\label{tab:total}.
\end{table}

\subsection{Data Analysis}
The analysis consisted of coding the transcripts, which involves dividing the qualitative data into parts with coherent content and assigning codes to these different parts \cite{runeson2012}. The coding was conducted by applying the two main types of data analysis methods: generation of theory and confirmation of theory. The aim of theory generation methods is to find hypotheses from the data, whereas theory confirmation methods are used to find support for previously generated hypotheses \cite{seaman1999}. Our initial codes originated from the goals of the study, the research questions, and other related variables of interest. As the analysis progressed, a number of codes were added. These post-formed codes were found iteratively by identifying recurring themes in the data and finding text parts which could not be coded with any of the preformed codes \cite{seaman1999}. In total, 28 codes where used, 16 of them was preformed. The coding was performed by the first and third author. The first interview was coded separately by both authors and combined into a resulting coded transcript. Because of the similarities between the selected codes of each author, the consequent interviews were equally divided between the authors instead. After a transcript had been coded by one author, it was validated by the other. If there were any disagreements regarding the assignment of codes, the particular text part was immediately discussed to agree on a final selection of codes.

\subsection{Threats to validity}
The threats to validity are outlined in this section. The threats to validity are discussed from an empirical validity point of view, which involves construct validity, internal validity, external validity, and reliability \cite{easterbrook2008,runeson2012}.

A threat to the \textbf{construct validity} is that the interview questions are not interpreted by the interviewees as the interviewers intended. This is addressed by commencing each interview with asking the interviewee to define the concepts of UX and IoT, respectively. When summarizing the answers, we use these definitions, together with the role of the interviewee, to judge from what point of view the development process is described. An additional threat to the construct validity is the semi-structure of the interviews. In some cases, an open question was used where it may have been preferable to use a closed question. For example, the question about UX activities was an open question and we did not ask the interviewee to list any activities in particular. 

In our case, the \textbf{internal validity} foremost regards the interview situation. One such threat is that the interviewee’s personal opinions may not represent that of the company. In that way, the answers could possibly be more related to the person rather than the company. Moreover, the interviewee’s role can be assumed to highly influence the answers and how much the interviewee knows about the subject in question. This threat is smaller for the small companies but for the larger companies, the threat cannot be ignored. For example, Interviewee F was not informed about  details of the company's UX work. During the interviews, we may have been more inclined to ask follow-up questions when the interviewees gave an answer which confirmed our theories, possibly resulting in confirmation bias. However, none of the persons who performed the interviews had any previous dealings with any of the companies or other relationships with them. In combination with a literature study, we consider the confirmation bias threat to be small. 

\textbf{The external validity} regards the aspect of the extent to which the results are generalizable to companies not part of the study. We interviewed both consultancy companies as well as product companies. In addition, we interviewed both young and old companies. However, only one larger product company was interviewed. The results are thus based mostly on consulting companies and start-ups. Hence, we cannot ignore the threats to validity. However, as argued by Flyvbjerg, the threats to generalizability should not be exaggerated \cite{flyvbjerg}.

One threat to the \textbf{reliability} concerns the coding. When the results were to be compiled from the tabulated and coded transcripts, we discovered that the way that the codes had been defined were too general. Furthermore, all interviews, except from the two at Company C and G were held in Swedish. We consider it as a threat to the reliability that information and meaning can be lost or changed due to the translation of quotes from Swedish to English. The translation has, to limited extent, also involved rephrasing and shorting some of the quotes, which may also contribute to this threat. 

\section{Results}\label{sec:results} 
This section summarizes the results from the 12 interviews performed with interviewees from 11 different companies. The following sub sections elaborate the results for each of the research questions presented in Section~\ref{sec:intro}.

\subsection{RQ1 How are UX requirements elicited in the context of IoT development in general?}
All participating companies state that they apply agile or iterative development methods. The consulting companies (Company A, B, C, E, F and G) are similar in the way that they are dependent on their customers' desires and it is generally the customer who directs how rigorous the requirements are specified. However, there are differences among the consultancy companies. In the initial part of the process, Company A and E focus on specifying mostly functional requirements, while Company B, C and G instead concentrate on exploring the underlying problem and origin of the customer's idea. Since company F is a large company with a separate UX department, Interviewee F could not describe their UX process in detail. The characterization of the development processes at the product companies varies from applying short iterations (Company D and J) or being directed by UX (Company J and K), to being unstructured and self-organizing (Company I and K). Apart from company A and E all companies describe their UX work as exploratory using for example prototypes and user stories instead of defining requirements. The development process at the innovation department at Company H is different to the process at Company H by being more iterative. Except from that Interviewee H explained that the innovative character of the development demands a more rapid process, the reason for applying a different development model \textit{" [It] is also that it's about Internet of Things. That is to say, it's unknown ground. The values are entirely untried"}. 

When describing the UX development process, the interviewees were asked if they apply any UX techniques. The techniques are presented in Table~\ref{tab:ux_activities}, categorized as either qualitative or quantitative. Extensive user research is foremost described by Interviewee G, H, J and K. Identifying the user groups and the underlying problem are seen as important. When asked how their UX decisions are made, Interviewee G answers \textit{"Research! [...] Both market research and then concept testing, basically"}. Interviewee J, describes that they have focused on the underlying needs rather than the product itself. Interviewee B, G, H, J, and K emphasize the importance of involving the end-users during the development process. For example, Company J have had beta testers, that provided both qualitative feedback and analytics data. Interviewee H sees it as one of their main activities during the process to go out in the field and talk to the end-users. Interviewee I bring up that they have had people testing their product using virtual reality. Even though it primarily was a marketing event, Interviewee I mention that they received valuable suggestions during that activity. Both interviewees from Company E believe that involving the end-user would be beneficial for their development process. However, it is rarely done. Interviewee E1 mentions that \textit{"In some cases, it may be that you may have to run some user test to test a hypothesis. But usually, it's enough to use our knowledge, i.e. previous experiences or [...] e.g. design guidelines."}. 

Prototyping is also something that is emphasized. Interviewee D stresses the use of 3D printing in order to be able to test different use cases early on. Interviewee B argues the use of easy and quick prototyping. However, the interviewee sees a problem with proceeding to generating solutions too quickly, since this involves a risk of losing the underlying meaning. Interviewee D experiences that it is easier to discuss a prototype than requirements, because \textit{"if you take [the prototype] to the developers, they exactly know what it's supposed to look like"}. 

\begin{table}[!hbt]
\centering
\includegraphics[scale=0.93]{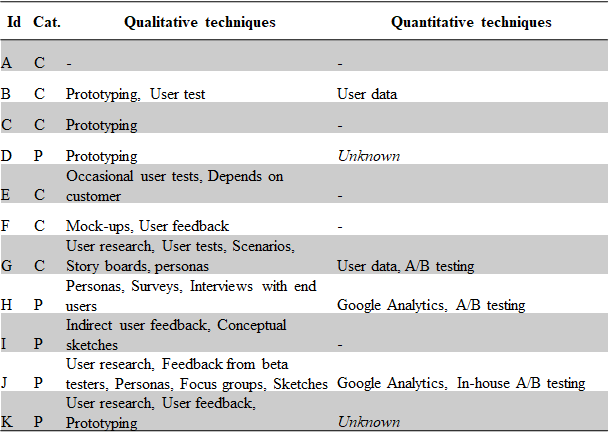} 
\caption{UX techniques during the development process, Category (Cat.): P = Product company, C = Consulting company}\label{tab:ux_activities}
\end{table}

To address RQ1, the handling of UX requirements are dependent on the customer's demands in the case of the consulting companies. However, company B, C and G tend to focus more on defining the problem together with the customer compared to company A and E that are more focused on requirements. When it comes to the product companies, UX requirements are generally not defined. Instead it is an ongoing process where for example user stories and prototypes are used to direct development.  

\subsection{RQ2 How are data-driven methodologies specifically utilized for IoT development to elicit UX requirements?} 

Companies that work in a more data-driven way do not generally see the product as finished when it has been released to market. For example, Interviewee K said \textit{"For us, it's not binary. It's not the traditional business mindset that you develop a product for a long time and then you release it and everyone will have access to it at the same time."}. 

Among the product companies, Company D, J, and K have all released their products to the market. None of them considers their product as finished and they explicitly describe that they use quantitative data from the product to develop the product also after-market release. Company D updates their products with new features and also collects data and statistics from the devices. Interviewee D emphasizes that update and data collection are important to their development and strategy; \textit{"In fact, all data that comes there can be used to create a better product"}. Company J uses Google Analytics data for various purposes, such as finding bugs, determining which functions that are used the most, and evaluating the set-up time. Information that comes from Google Analytics is seen as either a warning of that something is wrong or a sign of approval that it works as expected. However, Interviewee J claims that they are, to some extent, immature when it comes to using the data. The interviewee sees future possibilities with collecting other data than just which features that are used. One such possibility could be to extend the studying of behavioral data. In addition, the company is interested in behavioral data that concerns the physical product and not only the software. 

When it comes to A/B testing, Company J does it during the development process but not after-market release. Interviewee J means that the reason for not applying A/B testing in the field is that they \textit{"don't have that many customers yet. So we dare not risk that one particular solution may be bad"}. Company K develops new features that are released to a limited number of users. Interviewee K described that when \textit{"The product is out, it's already in thousands of homes. And we can do such a thing as doing a new feature, deploy it to a hundred users, and see if they are using it or if we want to do something more."}. 

Most interviewees that say that their company uses metrics related to UX, also argue that the quantitative data can be problematic and need to be complemented with data from, for instance, user tests or feedback from users. For example, as Interviewee G mean that numbers can be used to tell that something is important, but not why. Interviewee H stresses that \textit{"You have to use it with other data. You have to make interviews, and have contact with focus groups also to put it in context"}. Interviewee J sees Google Analytics data as an indication of that something is wrong; \textit{"It's usually just a catalyst, an indication that here's something strange."} Interviewee B mentions that they collect data in terms of different kind of feedback from users. In addition to working with Google Analytics, Company J also collects data from social media, support mail, and opinions from beta testers. Also, Interviewee D and K describe that they use customer feedback to improve the product. 

The consequences are that companies that apply data-driven methodologies (D, J, and K) are using the data as either confirmatory or as a warning that something is wrong. None of the companies let their UX design process be directed entirely by the quantitative data. 

\subsection{RQ3 Which are the challenges for UX and IoT?}
The interviewees identified some UX related challenges that are specific or more prominent when developing IoT compared to other systems. 

Interviewee D, E2, and J identify a challenge related to the IoT development process, which involves combining an agile software development process with hardware development. Interviewee J described this challenge as: \textit{"It's an obvious problem that, in a certain phase of the project, it's somewhat contradictory that [the software developers] want to wait as long as possible with deciding while the [hardware developers] must decide earlier"}.

Furthermore, Interviewee G argues that privacy and security is a UX challenge; \textit{"It's not necessarily a technical challenge, [it] is a UX challenge"}. Interviewee F stresses the connection between UX and security and means that an insecure device results in poor UX. 

Interviewee H means that UX for IoT can be seen as \textit{"an ecosystem of experiences"} and emphasizes that there is a number of factors that affects the experience that cannot be controlled but are affecting the UX. Something that is also mentioned by for example Interviewee D, is the problem of being dependent on other systems, such as the user's router or poor WiFi connection. Interviewee K argues that \textit{"The big challenges are when you have to build on systems that are not that good"}. Interviewee I and G sees it as an issue that it cannot be ensured that there is Internet connection available everywhere.

Interviewee H describes an interoperability issue as \textit{"One very basic thing is something that has been around for a long time, but is still difficult. And that is to connect things to each other."}. Interviewee G highlights the user perspective, which involves that the digitized products communicate with each other invisibly and consider it as a trap for IoT that the user does not have an intuitive perception towards that communication. Interviewee F regards it as problematic when different industries or even companies develop their own platforms and standards; \textit{"It will never work that each industry owns whole ecosystems. What is needed is openness and finding standards."}. In accordance, Interviewee G also describes the challenge of compatibility; \textit{"There's so many different solutions, applications out there [...]. There's just no standard"}. 

Interviewee D believes that it is easy to make too advanced services and that the installation needs to be simple.  This was also stressed by for example Interviewee C1 who said that they \textit{"call it plug-and-play"}, and Interviewee A that argued: \textit{"Anyone should be able to [install it] by picking it out of the box and starting it"}.  

Summarizing the challenges, there are challenges that are related to the development process at each company, but there are also challenges related to requirements that is not always controllable by the company itself since it involves also other systems.

\section{Discussion and Future Work}\label{sec:discussion} 
Based on the analysis of the transcriptions and codes of the interviews, we identify three topics of special interest that affect UX in IoT design in companies: \textit{Adapting to the situation}, \textit{Proactive vs. reactive}, and the \textit{The system context}. They are elaborated in the three following sub-sections. We conclude the discussion with a perspective on future work.  

\subsection{Adapting to The Situation}
Company A and E tend to define requirements early in the process to a larger extent than the other companies. One explanation can be that both companies are relatively old and have a tradition in hardware development respective industrial automation. The focus is also mainly on functional requirements, which also may be due to their respective background. The consulting companies, on the other hand, with a strong innovation and design profile (B and G) tend to define the problem together with their customer and focus on the underlying problem rather than defining requirements. However, as Interviewee E2, D and J mentioned, hardware in agile processes can be difficult since it is both expensive and time consuming to make hardware changes late in the process. 

A majority of the interviewees seems to consider prototyping as a natural part of their process. The use of different software prototyping tools described by Interviewee D, H, J and K is in line with contemporary UX research \cite{DeHaan2015,Kranz2010}. The use of rapid prototyping \cite{Kranz2010} is favored by Interviewee B, who at the same time considers it a risk that prototyping can undermine user research activities; there is a risk of being too confident if focusing on prototyping which leads to neglecting user research. 

There are examples among the companies that indicate that their design processes are both iterative, prototype-based, user-centered, and exploratory which is in accordance with how de Haan consider the development process to change due to IoT \cite{DeHaan2015}. Drawn from our results, we cannot confirm that the companies design process and their way of handling of UX requirements are due to the fact that they are developing for IoT. We consider that it is more likely that factors such as type of product, degree of innovation, company organization, and age of the company plays a greater role than the fact that it is IoT.   

The interviewees in our study brought up that the quantitative usage data itself does not tell anything about the underlying reason. Therefore, the interviewees propose that the quantitative data should be used together with qualitative data in order to understand, for instance, why a feature is used or not. Which is similar to what is proposed by Holmstr\"{o}m-Olsson et al. \cite{Olsson2015}. 

The companies are more or less immature in the use of quantitative data, something that they are also well aware of. Among the companies that collects data, they do not always know how to use the data. There is also a common skepticism regarding how useful the quantitative data is. However, there is a hype around data-driven methodologies that possibly lead to companies are afraid of lag behind if not adopting the new techniques. Data-driven methodologies are likely not always suitable. For example, Company J is using A/B testing during the development process, but not as the product are released to the market. They do not want to employ A/B testing as their customer base is too small and products too new.

With the advent of data-driven techniques, there is a knowledge gap and at the same time a hype which result in that companies collect large amount of data but are not mature in their way to make use of it. We hypothesize that there is a need for a better understanding of when and how a particular method or technique is appropriate to use to elicit, analyze and validate UX requirements.

\subsection{Proactive or Reactive?} 

Almost all of the companies apply some kind of prototyping whereas only a couple use data-driven approaches systematically. Company D, J, and K apply data-driven activities both during the development process and after the product is released to market. They have in common several of the preconditions for applying data-driven development suggested by Holmtr\"{o}m-Olsson et al. \cite{Olsson2012}. Firstly, none of the three companies consider their product as finished. Secondly, they have a product released to the market that they automatically collect data from. Thirdly, they have an organization where UX, software development, and product management are closely integrated. 

The interviewees agree that the data is difficult to use without interpretation. When it comes to UX, the data is mostly used as either confirmatory or as a warning that something is wrong. We believe that IoT is relatively unexplored which requires more creativity and innovation since there is a fewer number of applications to copy or take inspiration from. There seems to be a connection to the maturity of the products and markets and whether there is an emphasis on creative and proactive techniques (such as story boards and user and market research) or confirmatory and reactive techniques (such as usage data and user tests). The former is utilized more in more immature products and markets. Similar to De Haan \cite{DeHaan2015}, who states reactive and data-driven approaches \textit{"may simply lead to the most average HCI design ever created"}, there is also a connection to how radical innovation is being deployed and how long the iterations are. Hence, we hypothesize that longer iterations with more radical innovation is less suited for data-driven approaches whereas incremental innovation in short iterations are more suited for data-driven approaches. Obviously, hardware development has by nature longer iterations and hence more reliant on proactive approaches.  

\subsection{The System Context} 
An IoT device is always part of a larger system, dependent on a network, sometimes referred to as an ecosystem. This network may be of varied quality and will therefore in turn affect the quality of the Internet connection of the device. Furthermore, the other parts of the system are often developed by other companies with different goals. As expressed by Interviewee D, this is something that is out of the company's control, but it will still affect the UX of their IoT device. If the device also depends on additional systems, such as other IoT devices, interoperability issues may arise. The lack of standardization is an example of such an issue. This is brought up by Interviewee F and G during their respective interviews. These factors that are outside each company's control are also discussed by \cite{Shin2017}, who argues that these affect the QoS and thus the QoE (but are not part of UX according to them).

When the development of an ecosystem requires different industries to collaborate, it is an obstacle that, as Interviewee F described, separate industries wants to own the ecosystem. A collaboration requires standardization, but presumably, the reverse relationship - that standardization requires collaboration - is also a premise. Even though the concept of developing different systems part of a larger ecosystem is not new, we believe it is still largely not appropriately addressed. An IoT ecosystem will likely be even more diverse and coming from more vendors which emphasizes this problem from an UX perspective. 

\subsection{Future Work} 
One of the major challenges to IoT specifically but all software development in general is how to handle UX requirements when the products are part of a larger system, with less standards and control. In essence, there need to be a flexibility and adaptability to an unknown usage context. Especially when addressing immature markets and perhaps with immature products, the compromise between radical incremental innovation in relation to the UX will be key to product success. To study this, we suggest combining studies of comparative domains as well as applied research together with IoT companies to in depth understand their challenges and potential solutions. Furthermore, this study was conducted in a relatively limited geographical area. It would therefore be beneficial to extend the study into including companies in different geographical areas.

The relationship between UX and the challenge of privacy and in the context of IoT is something that, to our knowledge, there is little research on. A study could focus on the question if a high security and privacy level can have a positive impact on UX when it comes to IoT, especially with an ecosystem perspective. 

One interesting question that arises in the context of data-driven development, is how this approach to the development process affect the creativity when it comes to UX. As the quantitative measures becomes increasingly popular, it would be interesting to investigate the benefits and drawbacks from a creativity and innovation perspective and when different types of techniques and methods are the most suited.

The interest in UX among the companies could be described by the increased user-centeredness described by de Haan \cite{DeHaan2015} and the general shift towards UX found by Ovad et al. \cite{Ovad2015} and is not necessarily due to the fact that the companies develop IoT. A narrower categorization could be done, e.g. by comparing companies that all develop consumer IoT products. As an example, Company A and D do not involve their end-users to the same extent as Company J, H and K, for which a reason might be that they are B2B and not B2C. It is likely that the type of product influences the design process, which would be preferable to also compare with non-IoT companies.

Designing IoT can be particularly challenging since it, in many cases, does not have a traditional UI \cite{resnick2013ubiquitous} and is highly interconnected with other products, systems and services which affects the users' perceptions of the experience of use. From a user experience perspective, the actual size of the IoT system is irrelevant, and thus, many of the UX requirements may therefore be independent of size, but this would need to be investigated further.

\section{Conclusion}\label{sec:conclusion} 
In this study, we interviewed 11 companies working with IoT. The main characterizing factors are the hardware-software dilemma, agile and iterative development, fast-changing markets and technology as well as new usage contexts and interaction modes. Even though many aspects of IoT are not new, when combined they pose unique challenges for the companies when handling UX requirements. We believe that there is a need to better understand when a specific method is suited to help companies adapt to the specific situation at hand. Furthermore, there is a compromise to be made between an upfront, proactive analysis principle and an analysis of usage in running software, in a reactive manner. Even though there are proponents of data-driven, reactive methods, it is not clear that it leads to the best innovation in all situations. Lastly, UX requirements in a larger system of loosely connected companies are not well understood. Hence, there is a need to improve UX requirements elicitation and analysis methods in this context.  

Based on our study, indications are that there is no single solution which works for all companies and situations. Hence, we firmly believe in empirical understanding of the context and supporting companies with their unique problems and tailoring solutions that work in practice. 

\bibliography{litt} 

\end{document}